\documentclass[prb]{revtex4}
\usepackage{graphicx}
\begin{document}
\newcommand{\pr}{\partial}
\newcommand{\rta}{\rightarrow}
\newcommand{\lta}{\leftarrow}
\newcommand{\ep}{\epsilon}
\newcommand{\ve}{\varepsilon}
\newcommand{\p}{\prime}
\newcommand{\wf}{\omega_f}

\title{Two-Temperature Model of non-equilibrium electron relaxation: A Review}
\author{Navinder Singh}
\email{navinder@iopb.res.in; navinder_stat@yahoo.co.in}
\affiliation{Institute of physics, Bhubaneswar-751005, India} 

\begin{abstract}
The present paper is a review of the phenomena related to non-equilibrium electron relaxation in bulk and nano-scale metallic samples. The workable Two-Temperature Model (TTM) based on Boltzmann-Bloch-Peierls (BBP) kinetic equation has been applied to study the ultra-fast(femto-second) electronic relaxation in various metallic systems. The advent of new ultra-fast (femto-second) laser technology and pump-probe spectroscopy has produced wealth of new results for micro and nano-scale electronic technology.  The aim of this paper is to clarify the TTM, conditions of its validity and non-validity, its modifications for nano-systems, to sum-up the progress, and to point out open problems in this field. We also give a phenomenological integro-differential equation for the kinetics of non-degenerate electrons that goes beyond the TTM.
\end{abstract}
\maketitle

\newline\\
``The first processes, therefore, in the effectual studies of the sciences, must be ones of simplification and reduction of the results of previous investigations to a form in which the mind can grasp them.''
							             --- J.C. MAXWELL
\section{Introduction and Brief History}
The Two-Temperature Model (TTM) describes a non-equilibrium state between electrons and the lattice. When an electric field is applied to a metallic sample or the sample is photoexcited by a laser pulse, a state of non-equilibrium arise between electrons and the lattice. Since the time required to establish equilibrium in the electron gas (degenerate Fermi distribution) is much less than the time required to establish equilibrium between the electrons and the lattice phonons, the metal can be considered as composed of interacting subsystems, one electrons and other phonons. The hot degenerate (thermalized) electron gas relaxes to the bath phonons with relatively slow electron-phonon interactions. This problem was solved by V. L. Ginzburg and V. P. Shabanskii (1955)\cite{ginz} in high temperature limit ($T \gg T_D$, where $T_D$ is the Debye temperature). The complete solution for arbitrary temperatures was given by Kaganov, Lifshitz and Tanatarov (1956)\cite{kag}. They considered the weakly coupled electron and phonon subsystems and ignored the band structure of the metal, this work   was further extended to the case of a metal exposed to ultrashort laser pulses\cite{ani}, and today known as TTM. It is a pioneering work, that provides the basic model for ultra-fast pump-probe laser studies in metals\cite{ees86,scho87,fa92,fann92,sun94,rog95,hir03}. The theory of thermal relaxation of electrons in metals was further extended by P. B. Allen\cite{allen87}. It has been proposed that TTM can be derived from D. N. Zubarev's non-equilibrium statistical operator (NSO) method\cite{zub,kuzemsky1} by systematic approximations, with this, one can obtain even more advanced results than TTM including the non-Markovian effects\cite{pri1}. NSO method is based on a generalization of Gibbs' method of equilibrium states to non-equilibrium states. Gibbs equilibrium statistical mechanics is based upon the deep connection between additive integrals of motion (most common is energy) and the invariance of phase space distribution function due to Liouvelle theorem. This idea has been generalized to the non-equilibrium states by D. N. Zubarev using the concept of local quasi-integrals of motion and N. N. Bogolyubov's idea of a hierarchy of relaxation times i.e., evolution of the system from far-from-equilibrium stage to kinetic stage and then to the hydrodynamic stage as time proceeds. In our context of non-equilibrium electron-phonon system these time scales are in pico-second domain. It is a hope that some one among the readers will accomplish this very ambitious task of obtaining TTM from NSO method.

For the case of bulk semiconductors, theory of electron relaxation was given by Sh. M. Kogan (1963)\cite{kogan63}, with the similar assumptions as in the case of metals. It was extended and refined by S. Das Sarma (1990)\cite{das} and M.W.C. Dharma-Wardana (1991)\cite{war} using non-equilibrium Green's functions with dynamical screening and hot phonon effect. Large amount of work has been done and going on in nano-scale materials, the advent of nano-technology and ultra-fast lasers has completely revolutionized the field. However, this review is devoted to metallic nano-systems only. In metallic nano-systems the work of P. M. Tomchuk\cite{tom} is quite important(see the review\cite{fedo} for more detail on various phenomena caused by hot electrons in island metal films). 

The aim of this paper is to clarify the TTM, conditions of its validity and non-validity, and {\em modified } TTM as applied to nanoscale metallic systems, starting from first principles, without phenomenologically introducing power transferred to surface phenons as done in previous works\cite{tom}. The second aim of this paper is to point out new open problems about non-equilibrium electron relaxation in metallic nano-systems(see last paragraph of section III B), and also in bulk systems (beyond TTM, section IV). We also give a simple stochastic model to describe relaxation phenomena in non-equilibrium, non-degenerate (non-boltzmann) electronic subsystem. The adjective ``non-Boltzmann'' signifies that the gas is in non-equilibrium condition even in the classical approximation. We emphasis both aspects (namely, modified TTM for nano-systems using first principles and beyond TTM i.e., a general stochastic model) of the relaxation problem.

The paper is planned as follows. In the next section the approximation of {\em metal as a TTM system} is introduced and clarified. Section III is devoted to electronic relaxation in metallic nano-systems where TTM is used with appropriate boundary constraints. In section IV, by pointing out the limitations of TTM, we introduce a generalized stochastic model for electron-phonon interaction. In section V, we will end the manuscript with brief conclusion.

Before we go into the details of TTM, we first carefully define our notation. There is a lot of notational ambiguity in this research field, notation varies from author to author. However, for the sake of a coherent presentation we fix our notation as tabulated in table I, also, units and typical values are given for a physical feeling about the quantities involved. The terms non-degenerate, non-Fermi-Dirac, or non-thermalized electron distribution has been used in the literature, all have the same meaning, and point to the same physical condition of the electrons, but here, we adopt the term ``non-thermalized electron distribution'' for the sake of a unified presentation.
\begin{table}
\caption{Notation used for important physical quantities}
\begin{ruledtabular}
\begin{tabular}{lcr}

The Physical quantity          &  Notation used   &  Typical value and units\cite{sun94}\\\hline\hline

Temperature of the hot thermalized electron distribution &  $T_e$ & $\simeq 400$ K\\\hline

Temperature of the phonon bath  &  $T$   &  $ \simeq 300$ K \\\hline

Electronic heat capacity  &    $C_e =\gamma T_e$      & $\gamma = 66$ $J m^{-3}K^{-2}$ \\\hline

Phononic heat capacity &  $C_p$   & $\simeq 100~C_e$ \\\hline

Electron-phonon interaction constant &  $U_b$  &  $\simeq 10^{-19} ~ Joule$\\\hline

Electron-phonon interaction coefficient &  G  &  $\simeq 5\times10^{16}$ $J m^{-3} Sec^{-1} K^{-1}$ \\\hline

Decay time of non-thermal electron distribution & $\tau_{e-e}$ & $\sim 300$ femto-seconds (fs)\\\hline

Rise time of thermalized electron distribution & $\tau_{th}$ & $\sim 500$ fs \\\hline

Decay time of thermalized electron distribution & $\tau_{e-p}$ & $\sim 1-2$ peco-seconds (ps)\\

\end{tabular}
\end{ruledtabular}
\end{table}
\section{What is Two-Temperature Model(TTM)?}

Consider a metallic sample photoexcited by a femto-second laser pulse, because of the large difference between the electronic ($C_e$) and lattice(phononic $C_p$) heat capacities (with $C_p \gg C_e$ at room temperature), the femto-second laser pulse creates non-equilibrium electron distribution, leaving the lattice temperature essentially unchanged $T\simeq 300 K$. Then, over a time scale of hundreds of femto-seconds, the  non-equilibrium electrons redistribute their energies among themselves through electron-electron 
coulombic interaction, and return to a local equilibrium (among themselves) at a somewhat elevated temperature $T_e > T$. It is called the thermalized electron distribution.
This excited thermalized electron gas then cools(relaxes) via the electron-phonon interactions, giving up the excess energy to the phonon bath. Thus, the widely separated time-scales(the intra-electron and the intra-phonon relaxation times $\ll$ the inter-electron-phonon time scale) justifies defining the two temperatures $T_e$ and $T$. This motivates the Two-Temperature model\cite{kag,ani}. The TTM  describe this relaxation process, and has been  used extensively by the workers in the field of ultra-fast laser spectroscopy in nano-scale materials\cite{arb}. Briefly, the TTM assumes

(a) The electron-electron(coulombic) and the phonon-phonon(anharmonic) processes are much faster than the electron-phonon processes, so as to define $T_e$ and $T(\neq T_e~{\rm in~general})$ and maintain their local equilibrium distributions giving
\begin{eqnarray}
&&({\rm for\;\; electrons})~N_k = \frac{1}{e^{\beta_e(\ve-\ve_0)} +1}\;\;\;, \beta_e =\frac{1}{k_B
 T_e}\nonumber\\
&&({\rm for\;\; phonons})~N_f = \frac{1}{e^{\beta\hbar\omega_f} - 1}\;\;\;, \beta =\frac{1}{k_B T},
\label{1}
\end{eqnarray}
with a fermionic electron distribution at temperature $T_e$ and a bosonic phonon distribution at temperature 
$T \;(T<T_e)$.

(b) Homogeneous excitation and no spatial diffusion.

(c) Delta-pulse laser excitation.

So, only the collsion term is important in the Boltzmann transport equation, which has two contributions; phonon generation or phonon absorption when  electron scatters from one state to another. For a general discussion of the applicability of the Boltzmann kinetic equation in quantum cases see the review article by A. L. Kuzemsky\cite{kuzemsky3}. The transition rate (probability/second) of  an electron scattering from a state ${\bf k}$ to state ${\bf k^\p}$ with a phonon generation of energy $\hbar \wf$ 
is $W(k^\p|k)\delta(\ve_{k^{\prime}}-\ve_{k}-\hbar\omega)$, and for phonon absorption, transition rate is $W(k|k^\p)\delta(\ve_{k}-\ve_{k^\p}+\hbar\omega)$.
\begin{equation}
  W(k|k^\p) = \frac{\pi U_b^2}{\rho V S^2}\wf \;\;, \;\; {\bf f = k - k^\p }.
\label{1}
\end{equation}
With $U_b$ as the electron-phonon interaction constant (it appreas in the expression of time of flight of electrons and is a measure of interaction energy). It should be clearly distingueshed from electron-phonon interaction coefficient $G$ (see table I). ${\rho}$, V and $S$ is the metal density, unit cell volume and sound speed respectively.
So, the net phonon generation rate per unit volume is
\begin{equation}
\dot{N}_f = \int W(k|k^\p)\{(N_{f} +1)N_{k^{\prime}}(1-N_{k})-N_fN_k(1-N_{k^{\prime}})\}\delta(\ve_{k^{\prime}}-\ve_{k}-\hbar\omega)(2/(2\pi)^3)d\tau_{k^{\prime}}.
\label{2}
\end{equation}
This equation is known as Boltzmann-Bloch-Peierls (BBP) kinetic equation\cite{allen87}.
The delta function ensures the energy conservation. The factor of $2$ near the volume element is for electron spin degeneracy. Here $d\tau$ denotes k-space volume element, but later on $\tau$ is used to represent relaxation times, but confusion should not arise. One can show that the rate of energy transfer per unit volume by the electrons to bulk phonons is\cite{kag},
\begin{equation}
U_{bulk} = \int \dot{N}_f\hbar \wf V\frac{d\tau_f}{(2\pi)^3} = \left[\frac{m^2 U^2\omega^4 k_B}{2(2\pi)^3\hbar^3\rho S^4}\right][T_e - T].
\label{3}
\end{equation}
This can be cast in the following form\cite{ani}
\begin{eqnarray}
&&\frac{\partial (C_e T_e)}{\partial t} = -G (T_e - T), \\
&&\frac{\partial (C_p T)}{\partial t} = G (T_e - T).
\end{eqnarray}
The above coupled differential equations are the defining equations of the Two-Temperature Model (TTM) of hot electron cooling. $G$ is the electron-phonon interaction coefficient. If the system is not delta-pulse excited and heating of electron gas is going on in parallel with relaxation processes, then one can add a term $Q$ to the right hand side of equation (5) corresponding to the specific power absorbed by the sample from the laser field.

\section{Nano-systems: TTM and boundary constraints}

Nano-sized metallic systems are quite important for micro- and nano- electronic technology. In a typical electronic circuit the passage of current creates a situation of non-equilibrium between electrons and the phonons. In bulk metals, the electron energy relaxation is due to the cherenkov generation of acoustic waves. But this mechanism is not present in case of nano-sized metallic systems\cite{belo90}. TTM can not be directly applied to such small systems, because of the non-resonant nature of interaction between bulk phonons and the electrons. In nano-systems the main channel of electron energy transfer is the electron surface-phonon interaction(effective electron mean-free path becomes of the order of particle size). In the following we assume that TTM holds good with extra geometric constraints due to nano-size.

\subsection{Nano-scale Metallic Films}

Here we extend the TTM to nano-scale metallic films by considering surface phonon generation. In nano-scale metallic films the electron-surface-phonon interaction is important because film thickness is of the order of electron mean free path. The state of electrons and phonons is described by equilibrium Fermi and Bose functions with different temperatures. The new feature one has to consider is the geometric constraint of surface phonons. We obtain expressions for the energy transfer rate from thermalized hot electrons to surface-phonons, which is order of magnitude less than that for the bulk. The whole process occurs at pico-second time scales. We consider the case of a homogeneously photoexcited nano-scale metal film, and strong damping of surface phonons, due to the strong coupling with the substrate on which film was developed. Thus, there is no surface standing modes at the film surface\cite{nav04}.

Consider a hot Fermi electron distribution at temperature $T_e$ and a surface  phonon(2-D) equilibrium distribution at temperature $T \;(T<T_e)$. Both thermalized distibutions are weakly interacting subsystems. The energy flows from hot degenerate electron distribution to phonon bath. In the following, we calculate, in line with Kaganov etal\cite{kag},the energy transfered per second per unit volume from the hot thermalized electron distribution to the relatively cold phonon gas. The equilibrium distributions for electrons and phonons are given by equation (1). Energy and momentum conservation conditions gives,
\begin{equation}
\begin{array}{cc}
\ve_{k^\p} - \ve_{k} = \hbar\omega \;\;,{\bf k^\p- k = f}\;\;,  k^\p_x -k_x = f_x \;,\\
\; k^\p_y - k_y = f_y \;\,,\,k^\prime_z=-k_z\;\;,\omega = s f \;,\\
\ve_{k^\p}=\frac{\hbar^2{k^\p}^2}{2m}\;,\;\frac{\hbar^2}{2m}(2 [{k^\p}_x f_x + {k^\p}_y f_y]- f^2)=\hbar sf\;, \\
\frac{\hbar^2}{2m}[{k^\prime_x}^2-(k^\prime_x-f_x)^2 + {k^\prime_y}^2 - 
(k^\prime_y-f_y)^2] = {\hbar}sf\,.
\end{array}
\label{6}
\end{equation}
On simplifying,
\begin{equation}
\frac{\hbar^2}{2m}[2k^{\prime}\sin{\theta}\cos(\phi-\phi^{\prime})-f] = {\hbar}s.
\label{7}
\end{equation}
Where ${\phi}$ is the angle between $k_x$-axis and plane of incidence. ${\phi^{\p}}$ is the angle 
between scattered phonon direction and $k_x$-axis, and ${\theta}$ between incident electron direction and $k_z$  direction as shown in Figure 1. The probability $W$ per unit time that the electron in a state with wave vector $ k^\prime $
will scatter to a state with wave vector $k$ by  emitting a phonon of wave vector $f$ is;
\begin{equation}
W(k^\prime-f;k^\prime)=\alpha\delta(\ve_{k^\p} - \ve_k - \hbar
\omega),\,\,\alpha=(\pi U_s^2/\rho V S_s^2).
\end{equation}
With $U_s$ as the electron surface-phonon interaction constant. Here ${\rho}$, V and $S_s$ is the metal density, unit cell volume and surface sound speed respectively. The change per unit time per unit volume in the number of surface-phonons with wave vector $f$ and energy $\hbar\omega$ is (Boltzmann-Bloch-Peierls kinetic equation);

\begin{equation}
\dot{N}_f = \int\alpha\omega_f\{(N_{f} +1)N_{k^{\prime}}(1-N_{k})-
N_fN_k(1-N_{k^{\prime}})\}\delta(\ve_{k^{\prime}}-\ve_{k}-\hbar\omega)
(2/(2\pi)^3)d\tau_{k^{\prime}}.
\label{8}
\end{equation}
Using the energy and momentum conservation equations, the delta function can be written as
\begin{equation}
\delta (\ve_{k^\prime} -\ve_k - \hbar\omega)=\frac{2 m}{\hbar^2 f}\delta 
[(2 k^\prime \sin \theta \cos (\phi-\phi^\prime)-f)-\frac{2 m s}{\hbar}],
\end{equation}
and for a metal, we have $(f\sim 10^9  m^{-1})\gg (\frac{2 m s}{\hbar}\sim 10^7  m^{-1}) $. So the  equation (10) is
\begin{eqnarray}
&&\dot{N}_f = \int\alpha\omega_{f}\{(N_{f} +1)N_{k^{\prime}}(1-N_{k})-
N_fN_k(1-N_{k^{\prime}})\}\delta(2k^{\prime}\sin{\theta}\cos(\phi-\phi^{\prime})-f)\nonumber\\&& 
\times\frac{4m{k^\prime}^2}{2\pi^2\hbar^2 f}\sin{\theta}d{\theta}d{\phi}dk^{\prime}.
\end{eqnarray}
\begin{eqnarray}
&&\dot{N}_f = \left[\frac{4 m\alpha s \hbar}{(2\pi\hbar)^3}\right] \int_{k_m}^
{\infty}[(N_f +1)N_{k^\prime}(1-N_k)- N_f N_k (1-N_{k^\prime})]{k^\prime}^2 dk^{\prime}
\nonumber\\ 
&& \times \int_0^{\pi/2}\sin \theta d\theta\int_0^{2\pi}\delta[2k^\prime\sin
\theta\cos(\phi-\phi^\prime)-f]d\phi.
\label{7}
\end{eqnarray}
The last integral in the above equation is
\begin{eqnarray}
&&\int_0^{2\pi}\delta[2k^\prime\sin\theta\cos(\phi-\phi^\prime)-f]d\phi=\frac{1}{|2k^
{\prime}\sin\theta|}\nonumber\\ &&\times\left\{\frac{1}{|\sin(\phi_1-\phi^\prime)|}\int_
0^{2\pi}\delta(\phi-\phi_1)d\phi + \frac{1}{|\sin(\phi_2-\phi^\prime)|}\int_0^{2\pi}\delta
(\phi_2-\phi^\prime)d\phi \right\}\nonumber\\&& = \frac{1}{k^\prime\sin\theta\sqrt(1-
f^2/(4{k^\prime}^2\sin^2\theta))},
\label{8}
\end{eqnarray}
inserting this in (13) we get
\begin{equation}
\dot{N}_f = \left[\frac{ m\alpha s }{(2\pi\hbar)^2}\right] \int_{k_m}^
{\infty}[(N_f +1)N_{k^\prime}(1-N_k)- N_f N_k (1-N_{k^\prime})]{k^\prime} dk^{\prime}.
\label{6.8}
\end{equation}
The above mentioned process will always happen as from energy and momentum conservation,
$\sin{\theta} \simeq f/2k^{\prime}$, which holds good in a metal as $f<k^\prime$.
By inserting for $N_e$ and $N_f$ in equation (10) we get
\begin{equation}
\dot{N}_f = \left[\frac{ m\alpha s }{(2\pi\hbar)^2}\right]\left(\frac{e^{\beta\hbar\wf}-
e^{\beta_e\hbar\wf}}{e^{\beta\hbar\wf} - 1}\right)\int_{k_m}^\infty\frac{e^{\beta_e(\ve_k^\p
 -\hbar\wf-\ve_0)}k^\p dk^\p}{(e^{\beta_e(\ve_k^\p-\ve_0)}+1)(e^{\beta_e(\ve_k^\p-\hbar\wf-
\ve_0)}+1)}
\end{equation}
Here, we will make an approximation to solve the integral in the above equation. The first approximation is that the phonon energy $\hbar\omega_f(\sim meV) \ll k_B T_e(\sim eV)$, the electron energy.
So, $\beta_e\hbar\wf\sim 0 $. With this, the integral in Eq.(16) is
\[\frac{m}{\beta_e\hbar^2}\left[\frac{1}{e^{f(k_m)} +1}\right]\;\;,\;f(k_m) = 
\beta_e\left[\frac{\hbar^2 k_m^2}{2m}-\ve_0\right].\]
As $|\frac{\beta_e\hbar^2 k_m^2}{2m}|\ll|\beta\ve_0|$, the quantity in the square brackets is 
order of unity. Finally, the integral in Eq.(16) is $\frac{m\wf}{\hbar}(\frac{1}{\beta_e\hbar\wf})\sim \frac{m\wf}{\hbar}(1/(e^{\beta_e\hbar\wf}-1)) $. With all this, Eq.(16) takes the form
\begin{equation}
\dot{N}_f = \left[\frac{ m^2\alpha s \wf}{(2\pi\hbar)^2\hbar}\right]\left(\frac{e^{\beta\hbar\wf}-
e^{\beta_e\hbar\wf}}{(e^{\beta\hbar\wf}-1)(e^{\beta_e\hbar\wf}-1)}\right).
\label{6.10}
\end{equation}
Here $\alpha$ is defined in equation (9). The energy transfered by the electrons to the surface-phonons per unit volume per unit time is
\begin{equation}
U_{surface} = \frac{a^2}{(2\pi)^2}\int_0^{f_{Ds}}\dot{N}_f\hbar\wf 2\pi f df,
\end{equation}
where $f_{Ds}$ and $a$ are the Debye wave vector for the surface phonons and lattice constant respectively. From Eq.(17) and Eq.(18) with relations $\omega_{Ds} = S_s f_{Ds} \;\;,\;\hbar \omega_{Ds} = k_B T_{Ds}$ and setting $x = \hbar\wf/k_BT_e$, we get
\begin{eqnarray}
&&U_{surface} = \left[\frac{\pi U_s^2 m^2}{(2\pi)^3\hbar^2\rho a S_s^3}\right]\left(\frac{k_BT_{Ds}}{\hbar}\right)^4\nonumber\\
&&\times\left[\left(\frac{T_e}{T_{Ds}}\right)^4\int_0^{T_{Ds}/T_e}\frac{x^3}{e^x-1}dx-
\left(\frac{T}{T_{Ds}}\right)^4\int_0^{T_{Ds}/T}\frac{x^3}{e^x-1}dx \right].
\end{eqnarray}
Here, $T_{Ds}$ is the surface Debye temperature. Equation (19) can be simplified in two special cases, first, for low electron and phonon temperatures as compared to Debye temperature, i.e., $T, T_e \ll T_{Ds}$ , Eq.(19) reduce to
\begin{equation}
U_{surface} = \left[\frac{\pi U_s^2 m^2}{(2\pi)^3\hbar^2\rho a S_s^3}\right]\left(\frac{k_BT_{Ds}}{\hbar}\right)^4\left[\frac{T_e^4 - T^4}{T_{Ds}^4}\right]\int_0^\infty\frac{x^3}{e^x-1}dx.
\end{equation}
{\bf An important point to be noted in the above equation is that the electron to phonon 
energy transfer rate depends upon $4^{th}$ power of electron and phonon temperatures as 
compared to the corresponding case in the bulk(there it is $5^{th}$ power of electron and 
phonon temperatures\cite{kag})}. In second special case, when $T_e\;,\;T\gg T_{Ds}$ , we get
\begin{equation}
U_{surface} = \left[\frac{\pi U_s^2 m^2}{3(2\pi)^3\hbar^2\rho a S_s^3}\right]\left(\frac{k_BT_{Ds}}{\hbar}\right)^4\left[\frac{T_e - T}{T_{Ds}}\right].
\label{6.14}
\end{equation}
The above equation (Eq.(21)) is the basics of what is called the TTM. The surface Debye temperature $T_{Ds} = \frac{h}{k_B} f_{Ds}$, for two acoustic modes per atom is given by
\begin{equation}
\frac{L^2}{(2\pi)^2}\int_0^{f_{Ds}} 2\pi fdf = 2 N_{surface},
\end{equation}
which gives $f_{Ds} = \sqrt{8\pi n^{2/3}}$ , $n$ is the number density per unit volume. Now, for the bulk case\cite{kag}
\begin{equation}
U_{bulk} = \left[\frac{m^2 U_b^2\omega_{Db}^4k_B}{2(2\pi)^3\hbar^3\rho S_b^4}\right][T_e - T].
\end{equation}
From  Eq.(21) and Eq.(23) we have
\begin{equation}
\frac{U_{surface}}{U_{bulk}} = \frac{2}{3}\pi(8\pi)^{3/2}\left[\frac{U_s}{U_b}\right]^2\frac{n S_b^4}{a \omega_{Db}^4}.
\label{6.17}
\end{equation}
For a gold metal film, assuming $U_s = U_b$, with $a = 4.1\times10^{-10} m\;,\;\rho = 19.3
\times10^{3} kg/m^3\;,\;n = 5.9\times10^{28} m^{-3}\;,\;T_D = 185\;K\;,\;\omega_D = 2.42
\times10^{13} rads/sec$, the above ratio is $\sim 0.088$ or about 9 percent.

In brief, we have treated here the problem of energy relaxation of photo excited degenerate electrons in a nano-scale metal film. The new feature one has to add is the geometric constraint of surface phonons which is responsible for reduced energy transfer rate in nano-scale metal films as compared to bulk metals. In this simple calculation, the effect of electron-phonon screening and quantum confinement effects due to nano size, are not taken into account.
\begin{figure}
\includegraphics[height = 5 cm,width = 8 cm]{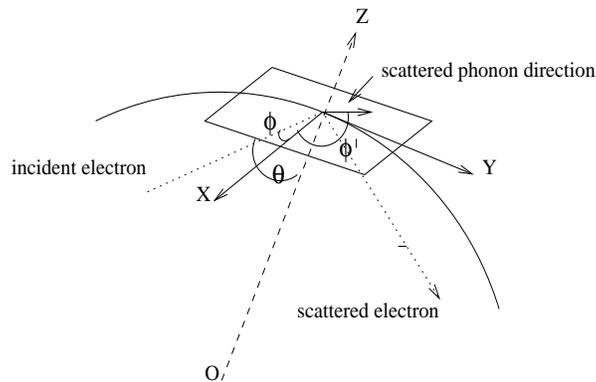}
\caption{Scattering of an electron from a surface.}
\end{figure}
\subsection{Hot electron relaxation in a metal nanoparticle: electron surface-phonon interaction}

As explained in the previous section,  TTM can not be directly applied to such small systems, because of the non-resonant nature of interaction between bulk phonons and the electrons. The main channel of electron energy transfer is the electron surface-phonon interaction(effective electron mean-free path becomes of the order of particle size)\cite{belo90}. P. M. Tomchuk and E. D. Belotskii give a quantum-kinetic treatment of hot electron energy relaxation in small metallic particles\cite{tom}. They phenomenologically introduced the power transferred from hot electrons to surface phonons (acoustic and capillary modes on the surface), and relate that to the microscopic parameters of the electron gas. In the following, we will give a calculation starting with BBP kinetic equation with geometric constraints, and obtain the expressions for electron surface-phonon coupling coefficient at both temperature limits. 

The reduced  dimensionality of nano-systems give rise to two important physical effects. One is quantum size effect, which give the quasi-continuous energy spectrum, and another is the geometric size effect(surface area/volume) which enhances the electron-surface interaction and heat diffusion to the bath. So, the following points are important:

(1)in quantum size regime when the particle size is typically less than 5nm, the band structure splits into discrete levels and the equilibrium partition
function of electrons will not be the same as that for the bulk. The function depends upon evenness or oddness of the number of electrons in the particle\cite{kubo};

(2) Since electronic mean free path(several hundred angstroms in metals) is more than the particle size, even at high temperatures, the scattering events from the surface of the particle will take place. If the time between two scattering events is less than the electronic internal thermalization time, one has to use non-equilibrium distribution functions to describe the problem of energy transfer from electrons to phonons\cite{cvo};

(3)In quasi-continuum regime(particle size more than 10nm)the main 
channel of electron energy loss is through electron surface interaction. One can use the Two-Temperature Model for electrons and the surface phonons. The only extra conditions to be imposed are of geometric nature. The present calculation is done in quasi-continuum regime, considering point (3). The points (1) and (2) are not included in the present calculation. It is 
assumed that TTM  holds good, but replacing
bulk phonons by surface phonons. This will not be applicable for small time 
scales when the electron distribution function is not Fermi-Dirac. It is to be noted that, as the hot electrons lose their energy to the lattice, and after some time, the lattice will become very hot, this heating will reduce drastically the electron mean free path and cause the failure of the applicability of the model. But for the case of metal particles it takes about 2 pico-seconds to transfer the energy to the lattice bath, 
so the present model is applicable within this time scale. The dispersion
relation used for the surface phonons is linear under Debye approximation 
and surface sound speed is determined in terms of elastic continuum theory assuming stress-free boundaries \cite{bal}.

We again consider the case of a homogeneously(no spatial diffusion) photo-excited metal nanoparticle. The equilibrium distribution functions of 
electrons and phonons are defined by equation (1).

We proceed on similar lines as in the previous section, and use the conservation of energy and momentum for one phonon scattering as shown in Figure 1. We use the BBP kinetic equation to calculate phonon generation rate. After imposing all boundary conditions we have
\begin{equation}
\dot{N}_f = \int\alpha\omega_f\{(N_{f} +1)N_{k^{\prime}}(1-N_{k})-
N_fN_k(1-N_{k^{\prime}})\}\delta(\ve_{k^{\prime}}-\ve_{k}-\hbar\omega)
(2V/(2\pi)^3)d\tau_{k^{\prime}}.
\label{518}
\end{equation}
\begin{equation}
\dot{N}_f = \left[\frac{ m^2\alpha s V\wf}{(2\pi\hbar)^2\hbar}\right]\left(\frac{e^{\beta\hbar\wf}-
e^{\beta_e\hbar\wf}}{(e^{\beta\hbar\wf}-1)(e^{\beta_e\hbar\wf}-1)}\right).
\label{6.19}
\end{equation}
Again $\alpha$ is defined in equation (9). Similarly, we calculate the energy transfered by electrons per unit volume per unit time to surface 
phonons. We use the elastic continuum theory assuming stress-free boundaries, for density of states on the particle surface\cite{bal}, Clearly
\begin{eqnarray}
&&U_{surface} = \int_{0}^{\omega_{Ds}}\dot{N}_f\hbar\wf (D_s(\wf)/((4/3)\pi R^3))d\wf\nonumber\\
&&D_s(\wf)=\frac{\wf R^2}{2 S_s^2},
\label{6.20}
\end{eqnarray}
Where ${\omega}_{Ds}$ is the Debye frequency for the surface phonons and $D_s(\omega)$ is
the surface phonon mode density\cite{bal}. Finally we obtain
\begin{equation}
U_{surface} = \eta_1\int_{0}^{\omega_{Ds}} 
{\wf}^3\left(\frac{e^{\beta\hbar\wf}-e^{{\beta}_e\hbar\wf}}{(e^{\beta\hbar\wf}-1)
(e^{{\beta}_e\hbar\wf}-1)}\right)d{\wf}.
\label{6.21}
\end{equation}
\[\eta_1 = \left[\frac{3\alpha m^2 V}{32{\pi}^3 \hbar^2 S_sR}\right]\]
The surface Debye frequency and the surface Debye temperature from
\begin{equation}
\int_0^{\omega_{Ds}} D_s(\wf)d\wf = 2(4\pi R^2a/V),
\label{6.22}
\end{equation}
are, $\omega_{Ds} = \sqrt{32 {\pi} a S_s^2/V}$ and $T_{Ds} = ({\hbar}S_s/k_B)\sqrt{32{\pi} a/V}$ 
respectively. With this we obtain
\begin{equation}
U_{surface} = \eta_1\left(\frac{k_BT_{Ds}}{\hbar}\right)^4
\left[\left(\frac{T_e}{T_{Ds}}\right)^4\int_0^{T_{Ds}/T_e}\frac{x^3}{e^x-1}dx-
\left(\frac{T}{T_{Ds}}\right)^4\int_0^{T_{Ds}/T}\frac{x^3}{e^x-1}dx \right].
\label{6.23}
\end{equation}
As before, the above equation (30) can be simplified in two special cases, first, for low electron and phonon temperatures as compared to Debye temperature, i.e., $T, T_e \ll T_{Ds}$ , we have
\begin{equation}
U_{surface} = \eta_1\left(\frac{k_BT_{Ds}}{\hbar}\right)^4\left[\frac{T_e^4 - T^4}{T_{Ds}^4}\right]\int_0^\infty\frac{x^3}{e^x-1}dx.
\label{6.24}
\end{equation}
Again, the point to be noted in the above equation is that the electron to phonon energy 
transfer rate depends upon $4^{th}$ power of electron and phonon temperatures as compared 
to the corresponding case in the bulk(there it is $5^{th}$ power of electron and phonon 
temperatures\cite{kag}). In second special case, when $T_e\;,\;T\gg T_{Ds}$ , we get

\begin{equation}
U_{surface} = \eta_1\left(\frac{k_BT_{Ds}}{\hbar}\right)^4\left[\frac{T_e - T}{T_{Ds}}\right].
\label{6.25}
\end{equation}
The important electron surface-phonon coupling coefficient (see equation (5)) in femtosecond pump-probe experiments 
for nanoparticles is
\begin{equation}
G = \left[\frac{3(\sqrt{32\pi}) m^2 U_s^2 k_B}{\pi{\hbar}^3\rho V } \right]\frac{1}{R}.
\label{6.26}
\end{equation}
For a gold nanoparticle of radius $R = 10 nm$, with $ U_s = 10^{-19} joule\;,\;a = 
4.1\times10^{-10} m\;,\;\rho = 19.3\times10^{3} kg/m^3\;,\;n = 5.9\times10^{28} m^{-3}\;,
\;T_D = 185\;K\;,\;\omega_D = 2.42\times10^{13} rads/sec$,
\begin{equation}
G \simeq 7.1\times 10^{13} joule\;\; m^{-3} sec^{-1} K^{-1}
\label{6.27}
\end{equation}
Which agrees with experiments\cite{gor} and P. M. Tomchuk's calculation. The electron-phonon coupling coefficient for the case of bulk is $\sim 5\times 10^{16}joule\;\; m^{-3} sec^{-1} K^{-1}$. So, $G$ (surface) is less by a factor of $10^{3}$ from that of bulk, which indicates suppression of electron energy transfer to phonons in case of nano-particles. The present results show that the electron surface-phonon coupling constant will increase with the reduction of the particle size. So the hot
electron thermalization time will reduce(fast relaxation) with decreasing size of the nanoparticle\cite{nisoli,stella}. The calculation does not include the effect of electron surface-phonon 
screening, but the fact that, due to electron wave function spill out and d- electron 
localization\cite{arb} in nanoparticles, the screening will be comparatively less as 
compared with the bulk. The question regarding the weight of the two factors, namely, 
surface to volume ratio (geometric factor), and reduction of electron phonon screening, 
in the thermalization of hot electron distribution is still open.
\section{Beyond Two-Temperature Model}

As discussed in section II, TTM assumes ``instantaneous'' thermalization of non-thermal electron distribution to a hot Fermi distribution, and then this hot Fermi distribution cools through electron-phonon interactions with the phonon subsystem. The assumption of ``instantaneous thermalization'' or ``adiabatic assumption'' is, in general, not in agreement with experimental observations\cite{fa92,fann92,sun94}, when one probe the system with very short and multi-wavelength laser pulses $(\sim 100 fs)$. Thus the TTM is clearly inadequate to account for non-thermal electron distributions. First phenomenological model to account for non-thermal electron distribution is given by Sun etal\cite{sun94}, which is quite successful in explaining  qualitatively the main features of the relaxation mechanism, and it is in agreement with experimental observations (typical pump-probe experiments) and numerical solutions of the Boltzmann equation. The model is applicable in the perturbative regime, where the system's response is linear and the measured changes in the reflectivity of the sample (thin Gold films of thickness $\sim 100 \AA$) can be related to the changes in electron distribution. The main idea is that the transient reflection or transmission of the sample is proportional to the real and imaginary parts of the dielectric constant, which is further related to the changes in the electron distribution--through joint density of states--due to a model developed by Rosei etal\cite{rosei}. Thus, in the linear perturbative regime, by knowing about transient reflectivity or transmissivity of the sample one can study the changes in the electron distribution. The rate equation model developed by them consists of dividing the whole sample into three interacting subsystems (see figure 2) (1) a very-low-density non-thermalized electron distribution, (2) high-density thermalized (Fermi) distribution, and (3) the phonon subsystem. Neglecting the particle exchange, the three subsystems interact according to:
\begin{eqnarray}
&& \frac{\pr N}{\pr t} = -\lambda_1 N -\lambda_2 N,\nonumber\\
&&C_e\frac{\pr T_e}{\pr t} = -G(T_e-T) + \lambda_1 N,\nonumber\\
&&C_p\frac{\pr T_l}{\pr t}=G(T_e-T)+\lambda_2 N.
\end{eqnarray}
Here, $N$ stands for the energy density stored in non-thermal part (subsystem 1), $C_e$ and $C_p$ are the electronic and lattice heat capacities, $T_e$ and $T$ are corresponding temperatures, $\lambda_1$ is the energy loss rate from non-thermal part to the thermalized part of the electronic part and $\lambda_2=\frac{G}{C_e}$ is the loss rate to the phonon bath. Thus the energy flows $N \propto e^{-t/\tau_{e-e}}$ from non-thermal part to the thermal part (subsystem 2) and also to the phonon bath (subsystem 3). In this process the temperature of the thermal part (subsystem 2) increases with time constant $\tau_{th}$ and decays (due the electron-phonon interaction) with time constant $\tau_{e-p}$ i.e., $\Delta T_e \propto (1-e^{-t/\tau_{th}}) e^{-t/\tau_{e-p}}$, clearly, we have $\tau_{th} > \tau_{e-e}$ (two loss channels from subsystem 1 and only one loss channel from subsystem 2). See table I for typical values of these time scales.
\begin{figure}
\includegraphics[height = 5cm, width = 12 cm]{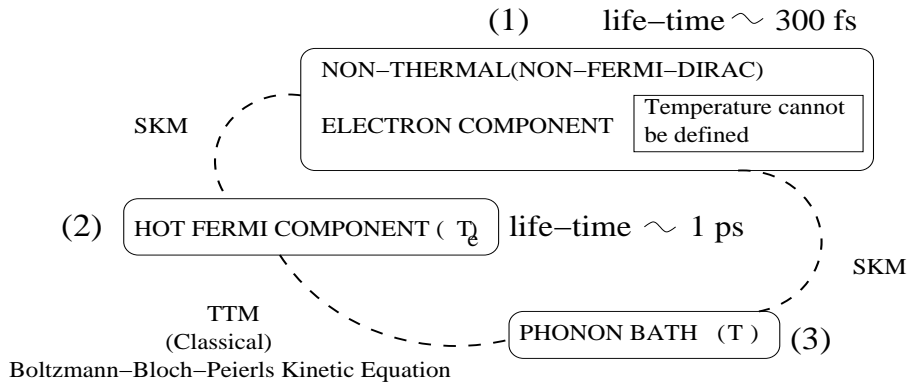}
\caption{Excited metal consists of interacting subsystems.}
\end{figure}

In the perturbative regime $(\Delta T_e \ll T_0 = ~ equilibrium ~ lattice~ temperature \simeq T)$, the contribution of the thermalized electron distribution to the transient reflectivity of the film is proportional to the electron temperature change, and contribution of the non-thermal part is proportional to the instantaneous energy density $N(t)$ of the non-thermal part. These two contributions to transient reflectivity depends upon the probe wavelength. This simple model yields remarkable agreement with the experimental results. However, a consistent micro-scopic model that take into account the general relationship between the optical response of the system and changes in electron distribution (not restricted in the perturbative regime) is not yet developed. The first principle model has to face the problem of non-equilibrium electron distributions, a part of the main problem of the non-equilibrium statistical mechanics itself. A possible way to the solution is to use D. N. Zubarev's non-equilibrium statistical operator method as mentioned in the introduction.
But micro-scopic models exists for experiments with four-wave mixing of pulses where 'spatial parametric effect'\cite{yajima} is important and one have to consider non-Markovian effects\cite{zubarev}.

To consider the non-equilibrium electron kinetics (non-thermal part), a stochastic kinetic model (SKM) is given below. By noting the fact that conduction electrons in metals can be treated free , an  analytical treatment is given which is based on a  generalization of the stochastic model known for a driven dissipative granular gas\cite{lev}. The generalized model is not applicable in an experiment where spatial parametric effect thus non-Markovian effects are important. The driven dissipative granular gas model is an interesting model where the particle-particle and the particle-bath collisions are parametrized in detail. More specifically , the total rate of collisions  suffered by a given ('tagged') particle is partitioned into the particle-bath collision  rate (fraction $f$) and the particle-particle collision rate (fraction $1-f$). Further, a fraction $\alpha$ of the total energy of the colliding particles is partitioned randomly between the colliding particles, while the remaining fraction $(1-\alpha)$ is dissipated through the frictional contact during the collisions. The system is kept in the dynamic (non-Boltzmannian) non-equilibrium condition by a constant drive. In our generalization to the electronic system, the bath has the obvious identification with phonons, and the drive is to be identified with the photo-excitation. Also, the possibly  dissipative electron-electron interaction has to be interpreted in terms of the coulomb interaction as screened by the dissipative polarization of the lattice. While, our generalization of the stochastic granular gas model to the electronic system covers time-dependent process relevant to the transient femtosecond photoexcitation, we have actually treated the steady state electron distribution under the cw(continuous wave) drive.

This stochastic generalized model describes the scattering events for time less than $\tau_{e-e}$ which could be quite large for the case of strong electron-phonon couplings. So, this model is quite general one. Also, for the case of continuous photoexcitation the total electronic system can be thought of divided into two subsystems, one non-degenerate and the other degenerate. This model captures the kinetic picture of the phenomenon in the non-degenerate subsystem.
\subsection*{The stochastic kinetic model (SKM) for non-thermal electrons}
Let $n_e(E)dE$ be the number of electrons lying in the energy range $\pm dE/2$ centred about $E$. The electron-electron 
collisions, assumed inelastic in general, are described by the process; $ E_i + E_i^\prime 
\longrightarrow E_f + E_f^\prime = \alpha (E_i + E_i^\prime)$ with $\alpha \leq 1$, in which the 
{\em tagged} electron  of energy $E_i$ collides with another electron of energy $E_i^\prime$ 
lying in the energy shell $E_i^\prime\pm\frac{1}{2}\Delta E_i^\prime$, and is scattered to the 
final state $E_f$. The scattering rate for this inelastic process is taken to be $(1-f)\Gamma 
n(E_i^\prime) dE_i^\prime$. Similarly, the electron-phonon scattering rate is given by $f\Gamma 
n_{ph}(E_i^\prime) dE_i^\prime$, with $n_{ph}(E_i^\prime) dE_i^\prime$ as the number of thermal 
 phonons in the phonon-energy shell $E_i^\prime\pm\frac{1}{2}\Delta E_i^\prime$. Here, the 
fraction $0\leq f\leq 1$ determines the relative strengths of the binary electron-electron and 
the electron-phonon collisions. Also, let the electrons are photoexcited at energy $E_{ex}$ above the Fermi energy at a rate $g_{ex}\delta(E-E_{ex})$, and then  recombine (deplete) from the non-degenerate distribution through recombination.  This depletion rate 
can be modelled by a term $-g_d\delta(E)n_e(0)$. Here, Fermi energy is set equal to zero for simplicity. The phonons are assumed to remain in thermal 
equilibrium at temperatures $T$. For non-degenerate electronic subsystem  energy to be the only label for the single particle states. The 
photo-excitation is taken to be homogenous over the sample, which is reasonable for a nanoscale metallic sample. For the above dissipative model driven far from equilibrium, the kinetics for the non-equilibrium electron number density $n_e(E)$ is given by the rate equation
\begin{eqnarray}
&&\frac{\partial n_e(E)}{\partial t}= -n_e(E)\int dE^\prime [n_e(E^\prime)(1-f)
+n_{ph}(E^\prime)f]\Gamma  \nonumber \\
&& +\int_0^1 dz p(z) \int dE^\prime \int dE^{\prime\prime} \delta(E-z\alpha (E^{\prime}+
E^{\prime\prime}))n_e(E^\prime)n_e(E^{\prime\prime})(1-f)\Gamma \nonumber \\
&&+\int_0^1 dz p(z) \int dE^\prime\int dE^{\prime\prime} \delta(E-z (E^{\prime}+
E^{\prime\prime}))n_e(E^{\prime})n_{ph}(E^{\prime\prime})f\Gamma \nonumber \\
&&+g_{ex}(t)\delta(E-E_{ex})-g_d\delta(E)n_e(0).
\end{eqnarray}

In the above, we have assumed the total energy $(E^{\prime}+E^{\prime\prime})$ for a binary collision to be partitioned such that a fraction $z$, with probability density $p(z)$, goes to the 
{\em tagged} electron of initial energy $E^{\prime}$, and $1-z$ to the colliding particle (electron
or phonon of initial energy $E^{\prime\prime}$). The inclusion of $\alpha$ in the electron-electron
collision takes care of the possibility of inelastic electron-electron collisions. Note that we
have suppressed the time argument ($t$) in the non-equilibrium electron-number density $n_e(E)$. 
Taking the energy Laplace transform
\begin{equation}
\tilde{n}_e(s)=\int_0^\infty e^{-sE}n_e(E)dE,
\end{equation}
of Eq.(36), we obtain,
\begin{eqnarray}
\frac{\partial}{\partial t} {\tilde{n}}(s) &=& - \Gamma \tilde{n}_{e}(s)[(1-f)N_e+fN_{ph}]+(1-f)
\Gamma 
\int_0^1 p(z) dz\,\,
\tilde{n}_{e}^2(\alpha z s)\nonumber \\
&&+f\Gamma \int_0^1 dz \,\,p(z)\,\,\tilde{n}_{e}(z s)\tilde{n}_{ph}(z s)\nonumber\\ && +g_{ex} (t)
 e^{-sE_{ex}}-g_d n_{e}(0).
\end{eqnarray} 

In the following, we will consider for simplicity the steady-state condition under constant (cw) 
photoexcitation, $g_{ex}(t)=g_{ex}$. A pulsed excitation can, of course, be considered in general.
Accordingly, we set $\frac{\partial}{\partial t} {\tilde {n}}_e (s) = 0$ above, and all
quantities on the R.H.S. of Eq.(38) are then independent of time.
In order to calculate the  steady-state electron distribution for the system in terms of the bath
(phonon) temperature and other rate parameters, we expand $\tilde{n}_e(s)$ in powers of the 
Laplace-transform parameter $s$ as
\begin{equation}
\tilde{n}_e(s)=N_e - s\langle E_e \rangle + s^2 \langle E_e^2 \rangle/2 . . . ,
\end{equation}
and equate the co-efficients of like powers of $s$. Thus, from the  zeroth power of s, we obtain 
at once
\begin{equation}
n_e(0) = (g_{ex}/g_d).
\end{equation}
Similarly, from the first power of s, we get,
\begin{equation}
\langle e_e \rangle=\frac{(f/2)\langle e_{ph}\rangle}{\rho_{e-ph}(1-\alpha)(1-f)+f/2}
+\frac{g_{ex}E_{ex}/\Gamma}{N_{ph}^2\rho_{e-ph}[\rho_{e-ph}(1-\alpha)(1-f)+f/2]}.
\end{equation}
\vspace{0.2cm}
In the above, we have taken a uniform limit for the energy partition: $p(z) = 1$. 

Here, we have defined $\langle e_e \rangle \equiv \langle E_e \rangle /N_e \equiv $
mean electron energy; $\langle e_{ph} \rangle \equiv \langle E_{ph} \rangle /N_{ph} 
\equiv $ mean phonon energy $(= k_B T_B)$; and $\rho_{e-ph}=N_e/N_{ph}\equiv$ electron-to-
phonon number ratio. It is to be noted that in the limit $\alpha = 1$ ({\it i.e.,} for elastic 
electron-electron collisions as is usually expected for an electronic system unlike the case of the
granular gas), and $g_{ex}=0$ ({\it i.e.,} no photo-excitation), we recover $\langle e_e 
 \rangle=\langle e_{ph}\rangle$,
{\it i.e.,} the electrons and the phonons are at the same temperature, as is physically expected 
under equilibrium conditions. In general, however, the mean electron energy in the steady state 
is not the same as the mean phonon energy, and the former depends on the excitation rate 
(the drive $g_{ex}$). 

The phenomenological integro-differential equation (Eq.38) is  a general result. 
Note that in (Eq.38), the probability $p(z)$ can be any `collision distribution' function. This kinetic equation can be solved by numerical simulations with specific choice of parameters. The important aspect of this stochastic model is that it describes the kinetics in non-degenerate electronic sub-system, because in this regime microscopic calculation for non-equilibrium distribution functions are complex, and no simple model is available. Our simple stochastic model gives a partial solution to the problem.

\section{Conclusion}

The present paper reviews the Two-Temperature Model based on Boltzmann-Bloch-Peierls kinetic equation. Its approximate validity is justified due to the fact that, in an excited metal, the intra-electron and intra-phonon relaxation time scales are much much less than the inter-electron-phonon relaxation time scales. In spite of the success of TTM in bulk case, it cannot be directly applied to spatially restricted systems, because of the non-resonant nature of the interaction between bulk phonons and the electrons. The main channel of energy relaxation is through electron surface-phonon interaction. Problems can be solved by considering surface effects, as explained in section III. We have seen (from the discussions in section IV) that TTM fails when one probe the sample with very short laser pulses (pulse width in the femto-second regime), and the non-thermal electron distribution has been experimentally observed. In this regard, we explained the rate-equation model of Sun etal and presented a stochastic kinetic model to describe the kinetics of non-thermal part of the electron distribution. So we  see that TTM is an approximation of BBP kinetic equation. In fact, BBP approach also has its own limited validity, it is valid for a rarefied electron gas and in the first Born approximation. Its success here lies in the fact that conduction electrons in metals can be treated free. For a general non-equilibrium electron-phonon system, non-equilibrium statistical operator method should give us more profound results, that will go beyond all the cases considered here.

\section{Acknowledgment}
I am grateful to Prof. N. Kumar and Prof. R. Srinivasan for helpful discussions and constant encouragement.

\end{document}